\documentclass[a4paper]{article}
\usepackage{multirow}
\usepackage{INTERSPEECH2021}
\usepackage{subfigure}

\title{Speech Disorder Classification Using \\ 
Extended Factorized Hierarchical Variational Auto-encoders}
\name{Jinzi Qi$^1$, Hugo Van hamme$^1$}

\address{
  $^1$Department Electrical Engineering-ESAT-PSI, KULeuven\\Kasteelpark Arenberg 10, Leuven, Belgium}
\email{jinzi.qi@kuleuven.be, hugo.vanhamme@kuleuven.be}

\begin{document}
\maketitle
\begin{abstract}
Objective speech disorder classification for speakers with communication difficulty is desirable for diagnosis and administering therapy. With the current state of speech technology, it is evident to propose neural networks for this application. But neural network model training is hampered by a lack of labeled disordered speech data. In this research, we apply an extended version of Factorized Hierarchical Variational Auto-encoders (FHVAE) for representation learning on disordered speech. The FHVAE model extracts both content-related and sequence-related latent variables from speech data, and we utilize the extracted variables to explore how disorder type information is represented in the latent variables. For better classification performance, the latent variables are aggregated at the word and sentence level. We show that an extension of the FHVAE model succeeds in the better disentanglement of the content-related and sequence-related related representations, but both representations are still required for best results on disorder type classification.

\end{abstract}

\section{Introduction}
Automated diagnosis of neurological diseases from the speech is receiving increased interest, as evidenced in the recent ADReSS-challenge \cite{luz2020alzheimer, luz2021detecting} which focuses on Alzheimer disease detection. Other neurological diseases such as Amyotrophic Lateral Sclerosis (ALS) and Parkinson's disease (PD) are reflected in speech disorders. Unfortunately, for many diseases, there is no cure but early detection and treatment may reduce complications, and progression may be slowed down with medication.

With the current state of the art in speech processing, it is a natural choice to explore neural network architectures to achieve accurate and objective automatic speech disorder classification.
In \cite{gillespie2017cross}\cite{kadiri2020parkinson}, classifications are done between control speakers and dysarthric speakers to achieve dysarthria detection. In \cite{gope2020raw}, convolutional neural networks and bidirectional long-short time memory network (BLSTM) are used for a binary classification between PD and ALS using raw speech. In \cite{kourkounakis2020detecting}, residual networks and BLSTMs are used for classifying different types of stuttering.

While neural networks are a sensible choice to classify speech disorders, they demand large amounts of labeled speech data to train the model. Disordered speech data, especially labeled data, are scarce due to the speaker's difficulty in speaking and the high cost to obtain reliable data labels. To avoid impact from insufficient data, authors have used auto-encoders for learning representations for disordered speech processing \cite{vachhani2017deep, yue2020autoencoder}. 
In this work we will apply an evolution of the basic auto-encoders, the Factorized Hierarchical Variational Auto-encoders (FHVAE), and extend it further.
An FHVAE \cite{hsu2017unsupervised, shon2018unsupervised} models the generative process of a sequence of segments in a hierarchical structure. It encodes a speech utterance into a segment-related variable (short time scale) and a sequence-related variable (long time scale) via two linked encoders. The segment-related latent variable represents the information that only appears in a single segment in the sequence, such as the phonetic content of the segment, and is conditioned on the sequence-related variable. The sequence-related variable reflects the features of the whole sequence, like the acoustic environment or speaker characteristics. The model hence offers a separation between speaker characteristics (sequence) and content (segment). 

The disentangled hierarchical representation of the FHVAE suggests using the sequence-related variable for inferring disorder type or severeness. However, we cannot expect the FHVAE model to disentangle speaker and content completely: dysarthric speakers exhibit a reduced vowel triangle~\cite{skodda2012impairment}, which would be reflected in the content variable. We therefore try to push this limitation in the sequence variable by expressing that speakers need to use the same content (segment) space, and hope to see that the disorder-related information is moved to the sequence space.  

We extend the FHVAE model with a regularization that will force the content (segment) space of different speakers to be similar at the expense of a different sequence (speaker) representation. We analyze to which extent we are capable of moving the disorder type information to the sequence level, where it belongs. To succeed in this goal, we will use a small corpus of phonetically labeled data. The training will hence be partly supervised.

This approach also opens perspectives for further work, e.g., automatic speech recognition or spoken language understanding for the disordered speech from the content variable. 

In this research, we take the first step towards this purpose, analyzing the FHVAE models on disordered speech and exploring how much disorder-related information the model can extract to the latent variables. 
We train the FHVAE model with labeled dysarthric speech data and use both the segment and sequence latent variable for disordered speech classification to quantify how much relevant information is found in each representation. 
Both latent variables are calculated for each segment, which is only 200~ms in this work, too short for reliable classification. To aggregate information over words or sentences, we compare simple averages or standard deviations or attention-based averages \cite{vaswani2017attention}, where we assume the attention mechanism can be trained to select those segments which show a high discrimination potential.

We introduce the basic and extended version of the FHVAE model and our classification scheme in section 2. In section 3, we describe the two databases we used and the experimental settings. Results and analysis will be provided in section 4, and section 5 gives conclusions.

\section{Model}

In this section, we introduce the original FHAVE~\cite{hsu2017unsupervised, shon2018unsupervised} and its proposed extension which we use to extract speech and speaker representations.
The classification network for disorder types and data aggregation strategies of the extracted latent variables will also be provided.

\subsection{Basic FHVAE}

\begin{figure}[t]
  \centering
  \includegraphics[width=\linewidth]{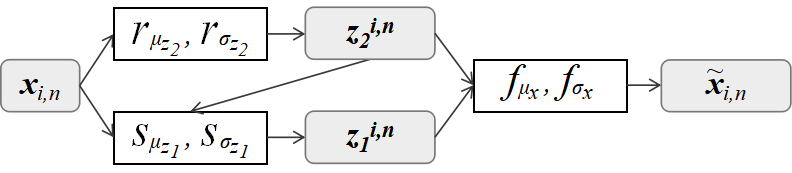}
  \caption{Dataflow in the basic FHVAE model.}
  \label{fig:flowchart}
\end{figure}

As a variant of classical VAE model, the FHVAE~\cite{hsu2017unsupervised, shon2018unsupervised} models a probabilistic hierarchical generative process of sequence data and extracts speech representations at both sequence and segment level. Suppose we have $I$ data sequences, each sequence $\textbf{X}_i$ containing $N$ segments $\textbf{x}_{i,n}$, $n = 1, 2, ..., N$. Each segment will be represented by two latent vectors $\textbf{z}^{i,n}_1$ and $\textbf{z}^{i,n}_2$, the first being conditioned on the second in the encoder (see Fig.~\ref{fig:flowchart}, eq. \ref{eq:qz1} and \ref{eq:qz2}). The generative model for segment $\textbf{x}_{i,n}$ needs an additional sequence-level variable $\boldsymbol{\mu}^i_2$, expressing that sequence information should be directed into $\textbf{z}^{i,n}_2$, while segment information should be found in $\textbf{z}^{i,n}_1$. Hence latent variables should follow these prior distributions:
\begin{equation}
    p(\textbf{z}^{i,n}_1) 
    = \mathcal{N}(\textbf{0}, \sigma^2_{\textbf{z}_1}\textbf{I})
\end{equation}
\begin{equation}
    p(\textbf{z}^{i,n}_2|\boldsymbol{\mu}^i_2) = \mathcal{N}(\boldsymbol{\mu}^i_2, \sigma^2_{\textbf{z}_2}\textbf{I}), \quad
    p(\boldsymbol{\mu}^i_2) 
    = \mathcal{N}(\textbf{0}, \sigma^2_{\boldsymbol{\mu}_2}\textbf{I})
\end{equation}
where $\sigma^2_{\textbf{z}_1},  \sigma^2_{\textbf{z}_2}$ and $ \sigma^2_{\boldsymbol{\mu}_2}$ are chosen upfront and $\textbf{z}^{i,n}_1, \textbf{z}^{i,n}_2$ are i.i.d samples. The prior enforces variables $\textbf{z}^{i,n}_2$ within the same sequence $i$ to be close. The generative model for the data is then:
% because in fig 1, i added a hat to reconstructed x 
% wait, I have an idea
%see your email,
% no. wait. It will mae the picture too dense

\begin{equation}
    p(\textbf{x}_{i,n}|\textbf{z}^{i,n}_1, \textbf{z}^{i,n}_2) 
    = \mathcal{N}(f_{\mu_x}(\textbf{z}^{i,n}_1, \textbf{z}^{i,n}_2), f_{\sigma_x}(\textbf{z}^{i,n}_1, \textbf{z}^{i,n}_2))
\end{equation}
where $f_{\mu_x}(\cdot)$ and $f_{\sigma_x}(\cdot)$ form the decoder ANNs of the FHVAE. This Gaussian mean serves as data reconstruction $\tilde{\textbf{x}}_{i,n}$ in Fig. \ref{fig:flowchart}.%good explanation

During inference, the latent variables $\textbf{z}^{i,n}_1, \textbf{z}^{i,n}_2$ are obtaining from segment $\textbf{x}_{i,n}$:
\begin{equation}
\label{eq:qz1}
    q(\textbf{z}^{i,n}_2|\textbf{x}_{i,n}) 
    = \mathcal{N}(r_{\mu_{\textbf{z}_2}}(\textbf{x}_{i,n}), r^2_{\sigma_{\textbf{z}_2}}(\textbf{x}_{i,n}))
\end{equation}
\begin{equation}
\label{eq:qz2}
    q(\textbf{z}^{i,n}_1|\textbf{x}_{i,n}, \textbf{z}^{i,n}_2) 
    = \mathcal{N}(s_{\mu_{\textbf{z}_1}}(\textbf{x}_{i,n}, \textbf{z}^{i,n}_2), s^2_{\sigma_{\textbf{z}_1}}(\textbf{x}_{i,n}, \textbf{z}^{i,n}_2))
\end{equation}
where $r_{\mu_{\textbf{z}_2}}(\cdot)$ and $r_{\sigma_{\textbf{z}_2}}(\cdot)$ are the ANNs encoding input $\textbf{x}$ to mean and variance of $\textbf{z}_2$,  $s_{\mu_{\textbf{z}_1}}(\cdot)$ and $s_{\sigma_{\textbf{z}_1}}(\cdot)$ are the ANN encoder of $\textbf{z}_1$. The sequence-level variable $\boldsymbol{\mu}^i_2$ can be inferred as:
\begin{equation}
\label{eq:mu2}
    \tilde{\mu}^i_2 = \frac{\sum^N_{n=1}r_{\mu_{\textbf{z}_2}}(\textbf{x}_{i,n})}{N+\frac{\sigma^2_{\textbf{z}_2}}{\sigma^2_{\boldsymbol{\mu}_2}}}
\end{equation}
 % of $\textbf{x}_n$. % HVH: what's the difference of \hat{z}_2 with r_{\mu_{z_2}} ?
% let me check the reference paper
%should be the same? I see in 
% Yes, i think so too. See https://arxiv.org/pdf/1709.07902.pdf p3 bottom vs eq (5).
% For the rest, I would not modify equations/notations. I understand it is overwhelming if you see it at first, but it should not be made simpler: the description will then loose the point.
%shall I change this part to r_{\mu_{z_2}}? Or leave it here, because we mentioned it's a posterior mean
% change to r_{\mu_{z_2}}
% done . Thanks! 
Then for the whole sequence $\textbf{X}_i$, the inference model is written as:
\begin{equation}
    q(\textbf{Z}^i_1, \textbf{Z}^i_2, \boldsymbol{\mu}^i_2|\textbf{X}_i) = q(\boldsymbol{\mu}^i_2|\textbf{x}_{i,n})
    \prod^N_{n=1}q(\textbf{z}^{i,n}_1|\textbf{x}_{i,n}, \textbf{z}^{i,n}_2)q(\textbf{z}^{i,n}_2|\textbf{x}_{i,n})
\end{equation}
where $\textbf{Z}^i_1 = \{ \textbf{z}_1^{i,n} \} ^N_{n=1}$, $\textbf{Z}^i_2 = \{ \textbf{z}_2^{i,n} \} ^N_{n=1}$.

The loss function of the basic FHVAE model is:
\begin{equation}
    LOSS = Loss_{LB} + Loss_{disc, z2}
\end{equation}
where $Loss_{LB}$ is the variational lower bound loss of the FHVAE model that guarantees the performance of the FHVAE model as an auto-encoder; $Loss_{disc, z2}$ is the discriminative loss of the sequence-related latent variable, which expresses that speaker identity should be inferrable from the sequence-related embedding vector, i.e., it is used to encourage separation in the latent space.

%A data flow chart of the FHVAE model is provided in Figure~\ref{fig:flowchart}.

\subsection{Extended FHVAE with regularization}

Speakers with dysarthria exhibit a reduced vowel triangle~\cite{skodda2012impairment}. If phonetic content is reflected in $\textbf{z}^{i,n}_1$, we would expect to find useful features for speech disorder type classification in both $\textbf{z}^{i,n}_1$ and $\textbf{z}^{i,n}_2$. 
We propose to regularize the $\textbf{z}^n_1$-space by expressing that latent representations of the same phoneme should be mapped near each other across sequences. This will require some labeled data, but the FHVAE training can be extended with unlabeled data by eliminating the regularization term for unlabeled sequences.

\begin{equation}
\begin{split}
    LOSS = &Loss_{LB} + Loss_{disc, z2} +\\ &Loss_{reg, z1} + Loss_{reg, z2}
\end{split}
\label{eq:etloss}
\end{equation}

In eq.~\ref{eq:etloss}, $Loss_{reg}$ is the regularization term loss of the latent variable which we added to encourage that the latent variables represent phonetic content ($\mathbf{z}_1$) or speaker properties ($\mathbf{z}_2$). These terms only contribute to the loss of labeled data.

To calculate the $Loss_{reg}$, two single layer classifiers are trained to predict the class label associated with $\textbf{z}^{i,n}_1$ and $\textbf{z}^{i,n}_2$ respectively.
The regularization loss is then the cross-entropy between the ground truth regularization label and the predicted label distribution. In this experiment, the regularization term used in $\textbf{z}_1$ is the phoneme classes at the center of the segment, and for $\textbf{z}_2$ it is the speaker identity.

\subsection{Disorder type classification network}
To explore speech disorder-related information extracted in latent variables $\textbf{z}_1$ and $\textbf{z}_2$, we use each variable as the input of a classification network. The classifier is a dense network with two layers: one layer with Relu activation and one output layer with softmax. 

Notice that since $\textbf{z}^{i,n}_1$ and $\textbf{z}^{i,n}_2$ are both segment-level latent variables, in our experiments, only containing extracted information from $200ms$ of speech, accurate classification could be difficult. 
Hence we aggregate latent variables within one word or sentence worth of speech data to achieve better performance. 
For $\mathbf{z}_2$, the mean (corrected for noise variance) eq.~\ref{eq:mu2} would be a candidate. For $\mathbf{z}_1$, the volume of the content space reached by an utterance is measured as the standard deviation of $\mathbf{z}_1$. For completeness, we add the mean and standard deviation of all latent variables.

Additionally, we also apply attention mechanisms~\cite{vaswani2017attention} to derive a merged embedding that focuses on the most discriminative segments. 
In the attention mechanism, 
we use $\mathbf{z}_1$ and/or $\mathbf{z}_2$ stacked over words or sentences as the value matrix, use a linear transformation of the value matrix as the key and train a fixed query matrix.
Except for $\textbf{z}_1$ and $\textbf{z}_2$, we also use the concatenated variable $\textbf{z}_{12}$.

\section{Experiments}
In the experiments, we use the Torgo~\cite{rudzicz2012torgo} English dysarthric speech database to train the FHVAE model. All data is labeled at the segment level; extensions with partly labeled data will be considered in the future. We work cross-domain and language because segment labels are not available for the Dutch evaluation data. Then latent variables $\textbf{z}_1$ and $\textbf{z}_2$ are extracted for the Dutch COPAS~\cite{copas} database of disordered speech and fed to the disorder classification network.

\subsection{The Torgo and COPAS databases}

The Torgo database \cite{rudzicz2012torgo} consists of aligned acoustics from 8 dysarthric speakers and 7 control speakers. The dysarthric speakers have cerebral palsy (CP) or amyotrophic lateral sclerosis (ALS), covering a wide range of intelligibility levels. All speakers were asked to read English text, designed as non-words, words or sentences. The phoneme transcriptions and alignments are generated using the free Wavesurfer tool~\cite{sjolander2000wavesurfer} according to the TIMIT phone set~\cite{timit}. In our experiments, we only use data from 11 speakers (7 dysarthric and 4 control speakers), which have phoneme-level alignments. We also cut off the silence segments in the audio according to the alignments. In total, 3463 audio files are used, and the number of phoneme classes considered during training is 49. 

The COPAS database \cite{copas} consists of recordings of Dutch Intelligibility Assessment (DIA)~\cite{dia} plus other materials. The DIA material is a designed list of 50 consonant-vowel-consonant words, and each DIA recording sentence contains 15-19 words. The database has samples from 319 speakers with or without speech disorder, and there are 7 pathology categories of speech disorder: dysarthria, voice disorders, cleft, articulation disorders, laryngectomy, glossectomy and impaired speech secondary to hearing impairment. In the database, some categories only have few speakers. Therefore, to guarantee minimal data sizes and to form a balanced dataset, we only use 5 types of speakers (control speakers, dysarthria, cleft, laryngectomy, impaired speech secondary to hearing impairment) and choose 29 speakers in each type. We cut each DIA recording into word-level audio pieces according to the provided word-level alignment. Because we only use in total 145 speakers' data of around 1.6 hours length, this can be insufficient for classifier training. So we augmented the data by changing the speed rate of the audio files. The rate speeds used are $0.8, 0.9, 1.0, 1.1, 1.2$, and the augmented data is 5 times the original dataset size, around 8 hours of speech.
% jq: I added a hat in equation 3, is it ok? See mail 
%email? ok thanks

\subsection{Experimental Setup}

Most hyperparameters are set to the values proposed by \cite{hsu2017unsupervised}. The speech features used in the FHVAE model are $80$ MEL filter bank energies with a frame advance of 10~ms. Each FHVAE segment is 20 frames, i.e., 200~ms of speech.  

In the FHAVE model, the encoder consists of 2 layers of bidirectional LSTM layers and 2 parallel dense layers for each latent variable. Each LSTM has 256 units, and the dimension of each latent variable is 32. The decoder consists of 2 bidirectional LSTM layers of the same size and 2 dense layers of 80 units for $\mu_x$ and $\sigma_x$ prediction, respectively. The classifier within the FHVAE model for each regularization term prediction is a single dense layer with softmax activation on the $11$ speaker identity classes ($\mathbf{z}_2$) or $49$ phone classes ($\mathbf{z}_1$).

During model training, we use around $11\%$ data for validation, $11\%$ for test and the rest for training.
Speakers have no overlap between each set. The training uses the Adam optimizer with a learning rate of 0.001, hierarchical sampling~\cite{hsu2018scalable}, a batch size of 500, 
an epoch limit of 500 with patience of 20 epochs. 
The chosen standard deviations for the priors 
$\sigma^2_{\boldsymbol{\mu}_2}, \sigma^2_{\textbf{z}_1}, \sigma^2_{\textbf{z}_2}$ are 
$1.0, 1.0, 0.5$.
Unless indicated otherwise, the weights of the discriminative loss of $\textbf{z}_2$ and the regularisation loss of $\textbf{z}_1$ and $\textbf{z}_2$ are $10, 1000, 10$ respectively.

For classification on the COPAS database, we run the experiments in a 6-fold cross-validation form. 
All speakers are partitioned into 6 groups, balanced for disorder type. A fold is then formed with 4 groups for training, one for validation and one for testing.
The data is shuffled before feeding into the classifier, and the averaged accuracy over the 6 experiments is used for evaluation. The classifier consists of a dense layer of 100 units with Relu activation followed by an output dense layer with softmax activation. 
Cross entropy is used as the training loss, and the Adam optimizer is used during training. Batch-size is 256, the maximal number of epochs is set to 500 with patience of 10 epochs.

\section{Results}

\subsection{Hyper parameters}

\begin{figure}[t]
  \centering
  \subfigure[]{
  \includegraphics[width=2.5in]{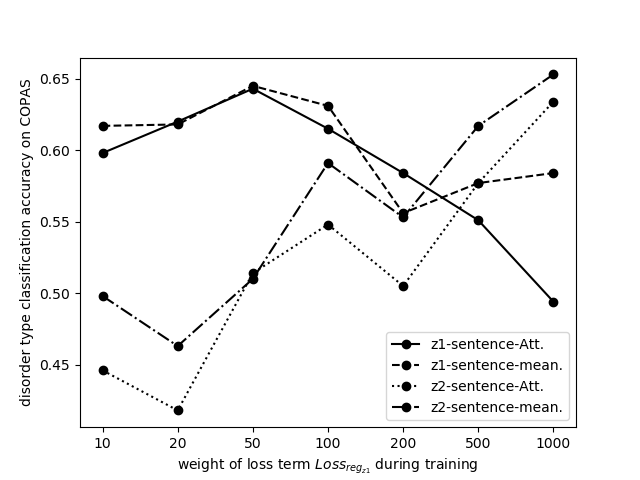}
  \label{fig:regacc}}
    \subfigure[]{
  \includegraphics[width=2.5in]{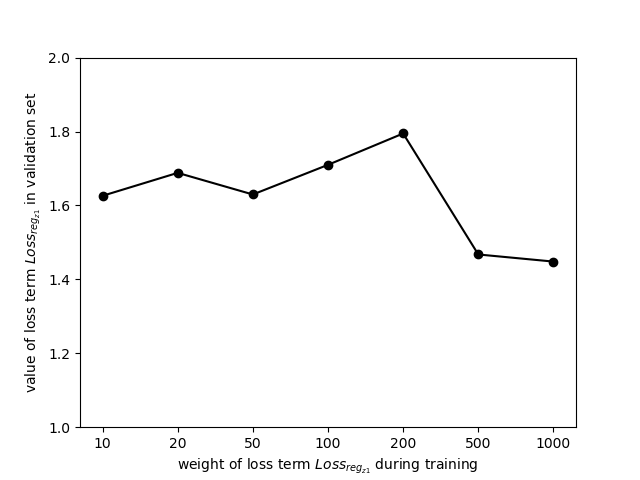}
  \label{fig:regloss}}
  \caption{Impact of the weight of loss term $Loss_{reg_{z1}}$ on (a) disorder type classification accuracy with sentence level $\textbf{z}_1$ or $\textbf{z}_2$ input, (b) value of loss term $Loss_{reg_{z1}}$ in the validation set.} 
  \label{fig:reglossweight}
\end{figure}

We first evaluate to which extent the FHVAE extension is effective in rearranging the latent spaces to contain disorder-related information. We vary the weight of the regularization weight in the segment-related $\mathbf{z}_1$ variable and observe its effect on the disorder type classification from the $\mathbf{z}_1$ resp. $\mathbf{z}_2$ latent variables in Fig.~\ref{fig:regacc} for sentence-level aggregation using unweighted mean or attention-based mean. 

We observe that the regularization weight indeed succeeds in moving disorder-related information from $\mathbf{z}_1$ into $\mathbf{z}_2$. With small weight, most of the disorder-related information is actually in $\mathbf{z}_1$, i.e., the basic FHVAE fails in building a disorder-independent latent content space. An invariant space would be a great asset for automatic speech recognition on dysarthric speech. The extended FHVAE does not succeed perfectly to disentangle content from disorder, though. 
In Fig.~\ref{fig:regloss}, we plot the regularization loss on the validation set as a function of regularization weight for the converged FHVAE model. Since the regularization loss is cross-entropy between true phone labels and prediction, this reflects classification accuracy on unseen data. We observe that phone classification based on the $\mathbf{z}_1$ variable becomes better with higher regularization weight, i.e., that we succeed in building a disorder-invariant content variable. We conclude that the regularization achieves its goal and select the maximal weight for further experiments.

\subsection{Disorder type classification}

We now explore the design choices for disorder representation.
To analyze the potential information related to disordered speech in the extracted variables, we use the segment-related latent variable $\textbf{z}_1$, sequence-related latent variable $\textbf{z}_2$ and the concatenated vector $\textbf{z}_{12}$ respectively as the classification network input. 
The baseline is using the mean vector of MFCC in each segment (dimension $80$) as the type classifier input.
We compare the classification performance when using variables of one segment directly and when aggregating the variables as the mean vector, standard deviation (std) vector or attention-based average on word or sentence level. 
Table\ref{tab:disordertype} shows classification accuracy when the above-mentioned choices are changed. Taking the 6-fold cross-validation into account, the reported accuracy is an average over 262800, 36000 and 2100 for segment, word and sentence-level classification, respectively. Since there are five balanced classes, chance level classification would have an accuracy of $0.20$.

\begin{table}[th]
  \caption{Speech disorder type classification accuracy at segment, word or sentence level from MFCCs or latent variables extracted with the basic or extended FHVAE.}
  \label{tab:disordertype}
  \centering
  \begin{tabular}{ l@{}l  c c c}
    \toprule
    \multicolumn{2}{c}{\multirow{2}{*}{\textbf{Classifier Input}}} & \multirow{2}{*}{MFCC} & \multicolumn{2}{c}{FHVAE Model} \\
    \cline{4-5}
    & && Basic & Extended \\
    \midrule
    MFCC &-1 segment                & 0.50 & - & - \\
      & -word-$Att.$                & 0.52 & - & - \\
      & -word-mean                  & 0.49 & - & - \\
      & -sentence-$Att.$            & 0.56 & - & - \\
      & -sentence-mean              & \textbf{0.60} & - & - \\
      & -sentence-std               & 0.57 & - & - \\
    $\textbf{z}_1$ & -1 segment     & - & 0.44 & 0.42 \\
      & -word-$Att.$                & - & 0.51 & 0.47 \\
      & -word-mean                  & - & 0.48 & 0.48 \\
      & -sentence-$Att.$            & - & 0.54 & 0.50 \\
      & -sentence-mean              & - & 0.60 & 0.59 \\
      & -sentence-std               & - & 0.55 & 0.55 \\
    $\textbf{z}_2$ & -1 segment     & - & 0.46 & 0.49 \\
      & -word-$Att.$                & - & 0.51 & 0.55 \\
      & -word-mean                  & - & 0.50 & 0.55 \\
      & -sentence-$Att.$            & - & 0.48 & 0.64 \\
      & -sentence-mean              & - & 0.55 & 0.65 \\   
    $\textbf{z}_{12}$& -1 segment   & - & 0.50 & 0.50 \\
      & -word-$Att.$                & - & 0.55 & 0.56 \\
      & -word-mean                  & - & 0.53 & 0.56 \\
      & -sentence-$Att.$            & - & 0.61 & 0.62 \\
      & -sentence-mean              & - & \textbf{0.62} & \textbf{0.67} \\
      & -sentence-mean\&std         & - & 0.61 & 0.64 \\
    \bottomrule
  \end{tabular}
\end{table}

In general, aggregating data over words and sentences results in better accuracy, as expected. Where we expected the standard deviation in $\mathbf{z}_1$ space to work well, because it reflects the speaker's coverage of the acoustic space, the shift in the acoustic space, represented as the mean of $\mathbf{z}_1$ features seems to work better. The attention mechanism never succeeds in performing better than the mean.
Comparing the classification accuracy with MFCC and latent variables from FHVAE models, the best FHVAE model shows an advantage in extracting speech disorder type information from speech. 
For the extended model, its variable $\textbf{z}_2$ gives better accuracy than $\textbf{z}_1$ and basic model variables. The concatenated variable $\textbf{z}_{12}$, which gathers information from both segment and sequence level, always shows accuracy exceeding that of any single variable.
The best classification performance we obtained hence still combines both latent variables, reaching a classification accuracy of $0.67$ from sentence-level information using the extended FHVAE. 
The accuracy is not perfect, but we do have a fair agreement between predicted and ground-truth disorder types, which shows the potential that the latent variable of the FHVAE model contains disorder-related information for patients.

\section{Conclusions}

 In this research, we investigated the potential of an FHVAE model for dysarthric speech processing tasks. Auto-encoders are appealing since we can train them unsupervisedly, and labeling dysarthric speech is hard. However, the purpose of an FHVAE is to disentangle the "who" and the "what" in speech. For dysarthric speech, it is unrealistic to hope that the "what" representation would not be affected by the disorder type, so we proposed an FHVAE extension to regularize the model on (partly) labeled data. We showed that this extension does succeed in disentangling the "who
" and the "what" better, in that the "what" (content or segment-related) representation shows better phone classification while the "who" (sequence-related) representation shows better disorder type classification.
However, both representations are not fully disentangled, and for the best classification results, both representations are required.

In future work, we will explore using additional unlabeled dysarthric speech data during the FHVAE training as well as using labeled normal speech data. Also, the FHVAE can be extended with phone-specific Gaussian priors in the segment-related latent space, which could strengthen the disentanglement. Finally, an evaluation of the extent to which the learned content-related speech representations benefit disorder-independent automatic speech recognition will be performed.

\section{Acknowledgements}
The research was supported by KUL grant CELSA/18/027 and the Flemish Government under “Onderzoeksprogramma AI Vlaanderen”.

\bibliographystyle{IEEEtran}

\bibliography{mybib}

\begin{thebibliography}{10}
\providecommand{\url}[1]{#1}
\csname url@samestyle\endcsname
\providecommand{\newblock}{\relax}
\providecommand{\bibinfo}[2]{#2}
\providecommand{\BIBentrySTDinterwordspacing}{\spaceskip=0pt\relax}
\providecommand{\BIBentryALTinterwordstretchfactor}{4}
\providecommand{\BIBentryALTinterwordspacing}{\spaceskip=\fontdimen2\font plus
\BIBentryALTinterwordstretchfactor\fontdimen3\font minus
  \fontdimen4\font\relax}
\providecommand{\BIBforeignlanguage}[2]{{%
\expandafter\ifx\csname l@#1\endcsname\relax
\typeout{** WARNING: IEEEtran.bst: No hyphenation pattern has been}%
\typeout{** loaded for the language `#1'. Using the pattern for}%
\typeout{** the default language instead.}%
\else
\language=\csname l@#1\endcsname
\fi
#2}}
\providecommand{\BIBdecl}{\relax}
\BIBdecl

\bibitem{luz2020alzheimer}
S.~Luz, F.~Haider, S.~de~la Fuente, D.~Fromm, and B.~MacWhinney, ``Alzheimer's
  dementia recognition through spontaneous speech: The adress challenge,''
  \emph{arXiv preprint arXiv:2004.06833}, 2020.

\bibitem{luz2021detecting}
S.~Luz, F.~Haider, S.~de~la Fuente, D.~Fromm, and B.~MacWhinney, ``Detecting
  cognitive decline using speech only: The adresso challenge,'' \emph{medRxiv},
  2021.

\bibitem{gillespie2017cross}
S.~Gillespie, Y.-Y. Logan, E.~Moore, J.~Laures-Gore, S.~Russell, and R.~Patel,
  ``Cross-database models for the classification of dysarthria presence.'' in
  \emph{Interspeech}, 2017, pp. 3127--3131.

\bibitem{kadiri2020parkinson}
S.~R. Kadiri, R.~Kethireddy, and P.~Alku, ``Parkinson’s disease detection
  from speech using single frequency filtering cepstral coefficients,'' in
  \emph{Interspeech}, 2020, pp. 4971--4975.

\bibitem{gope2020raw}
J.~Mallela, A.~Illa, Y.~Belur, N.~Atchayaram, R.~yadav, P.~Reddy, D.~Gope, and
  P.~K. Ghosh, ``Raw speech waveform based classification of patients with als,
  parkinson’s disease and healthy controls using cnn-blstm,'' in
  \emph{Interspeech}, 2020, pp. 4586--4590.

\bibitem{kourkounakis2020detecting}
T.~Kourkounakis, A.~Hajavi, and A.~Etemad, ``Detecting multiple speech
  disfluencies using a deep residual network with bidirectional long short-term
  memory,'' in \emph{IEEE International Conference on Acoustics, Speech and
  Signal Processing (ICASSP)}.\hskip 1em plus 0.5em minus 0.4em\relax IEEE,
  2020, pp. 6089--6093.

\bibitem{vachhani2017deep}
B.~Vachhani, C.~Bhat, B.~Das, and S.~K. Kopparapu, ``Deep autoencoder based
  speech features for improved dysarthric speech recognition.'' in
  \emph{Interspeech}, 2017, pp. 1854--1858.

\bibitem{yue2020autoencoder}
Z.~Yue, H.~Christensen, and J.~Barker, ``Autoencoder bottleneck features with
  multi-task optimisation for improved continuous dysarthric speech
  recognition,'' in \emph{Interspeech}, 2020, pp. 4581--4585.

\bibitem{hsu2017unsupervised}
W.-N. Hsu, Y.~Zhang, and J.~Glass, ``Unsupervised learning of disentangled and
  interpretable representations from sequential data,'' in \emph{Advances in
  Neural Information Processing Systems}, 2017, pp. 1878–--1889.

\bibitem{shon2018unsupervised}
S.~Shon, W.-N. Hsu, and J.~Glass, ``Unsupervised representation learning of
  speech for dialect identification,'' in \emph{IEEE Spoken Language Technology
  Workshop (SLT)}.\hskip 1em plus 0.5em minus 0.4em\relax IEEE, 2018, pp.
  105--111.

\bibitem{skodda2012impairment}
S.~Skodda, W.~Gr{\"o}nheit, and U.~Schlegel, ``Impairment of vowel articulation
  as a possible marker of disease progression in parkinson's disease,''
  \emph{PloS one}, vol.~7, no.~2, p. e32132, 2012.

\bibitem{vaswani2017attention}
A.~Vaswani, N.~Shazeer, N.~Parmar, J.~Uszkoreit, L.~Jones, A.~N. Gomez,
  L.~Kaiser, and I.~Polosukhin, ``Attention is all you need,'' in
  \emph{Advances in Neural Information Processing Systems}, 2017, pp.
  5998–--6008.

\bibitem{rudzicz2012torgo}
F.~Rudzicz, A.~K. Namasivayam, and T.~Wolff, ``The torgo database of acoustic
  and articulatory speech from speakers with dysarthria,'' \emph{Language
  Resources and Evaluation}, vol.~46, no.~4, pp. 523--541, 2012.

\bibitem{copas}
G.~Van~Nuffelen, M.~De~Bodt, C.~Middag, and J.-P. Martens, ``Dutch corpus of
  pathological and normal speech (copas),'' Antwerp University Hospital and
  Ghent University, Tech. Rep., 2009.

\bibitem{sjolander2000wavesurfer}
K.~Sj{\"o}lander and J.~Beskow, ``Wavesurfer-an open source speech tool,'' in
  \emph{Sixth International Conference on Spoken Language Processing}, 2000.

\bibitem{timit}
J.~Garofolo, L.~Lamel, W.~Fisher, J.~Fiscus, D.~Pallett, and et~al, ``Timit
  acoustic-phonetic continuous speech corpus ldc93s1,'' Web Download, 1993.

\bibitem{dia}
M.~De~Bodt, C.~Guns, and G.~Van~Nuffelen, ``Nsvo: Nederlandstalig
  spraakverstaanbaarheidsonderzoek,'' Herentals: Vlaamse Vereniging voor
  Logopedisten, 2006.

\bibitem{hsu2018scalable}
W.-N. Hsu and J.~Glass, ``Scalable factorized hierarchical variational
  autoencoder training,'' in \emph{Interspeech}, 2018, pp. 1462--1466.

\end{thebibliography}

\end{document}